# Re-analysing the phase sequence in $Eu_{0.8}Y_{0.2}MnO_3$

J. Agostinho Moreira, A. Almeida, W. S. Ferreira and M. R. Chaves

*IFIMUP and IN- Institute of Nanoscience and Nanotechnology. Departamento de Física da Faculdade de Ciências da Universidade do Porto. Rua do Campo Alegre, 687. 4169-007 Porto. Portugal.*

S. M. F. Vilela and P. B. Tavares

*Centro de Química. Universidade de Trás-os-Montes e Alto Douro. Apartado 1013, 5001-801. Vila Real. Portugal.*

*e-mail: jamoreir@fc.up.pt

## Abstract

To clarify the controversy concerning the ferroelectricity of the lower temperature magnetic phases of $Eu_{0.8}Y_{0.2}MnO_3$, hence its multiferroic character, we have studied in detail the temperature dependence of the electric polarization. The existence of a spontaneous polarization in 30K < T < 22K, provides clear evidence for the ferroelectric character of the re-entrant magnetic phase, stable in that temperature range. Both lower temperature magnetic phases allow easily-induced electric polarization, which actually mask any spontaneous polarization.

Two years ago, Hemberger et al[1] published a work about the structural, thermodynamic, magnetic and dielectric properties of the orthorhombic Y-doped EuMnO$_3$ system (Eu$_{1-x}$Y$_x$MnO$_3$, with $0 \leq x < 1$). Based on their experimental results and on theoretical arguments, these authors proposed a phase diagram $(x,T)$ for Eu$_{1-x}$Y$_x$MnO$_3$, in the $0 \leq x \leq 0.5$ range,[1] more detailed than the one proposed by Ivanov et al, the year before.[2] Among these compounds, Eu$_{0.8}$Y$_{0.2}$MnO$_3$ has been reported as the only magnetoelectric multiferroic system among the rare-earth perovskite manganites, because it exhibits both ferroelectricity and ferromagnetism in the same thermodynamic phase.[1] In the following, we shall summarize some of the main features of the phase sequence of Eu$_{0.8}$Y$_{0.2}$MnO$_3$ presented in Figure 8 of Ref. 1. The paramagnetic and paraelectric phase of Eu$_{0.8}$Y$_{0.2}$MnO$_3$ transforms into an antiferromagnetic phase (AFM-1) at T$_N$ = 48K, presumably with an incommensurate sinusoidal collinear arrangement of the Mn$^{3+}$ spins. The anomalies detected in both specific heat and electric permittivity have revealed another phase transition at T$_{AFM-2}$ = 30K. Double magnetic hysteresis loops at 25K were reported, revealing the antiferromagnetic character of the phase below T$_{AFM-2}$, hereafter called AFM-2.[1] Based on the anomalous behaviour observed in the electric permittivity and magnetization curves, a canted antiferromagnetic phase (AFM-3) below T$_{AFM-3}$=22K has been proposed.[1] According to Hemberger et al,[1] Eu$_{0.8}$Y$_{0.2}$MnO$_3$ becomes ferroelectric below T$_{AFM-2}$ = 30K. The electric polarization was obtained from the time integration of a pyroelectric current, after poling the sample under an electric field of 1 kV/cm.[1] The ferroelectric character of both low temperature AFM-2 and AFM-3 magnetic phases was also found by Valdés et al in Eu$_{0.75}$Y$_{0.25}$MnO$_3$.[3] Taking into account the weak ferromagnetic character of the AFM-3 phase, as well as the electric polarization below T$_{AFM-2}$ = 30K, Hemberger et al[1] have proposed a noncollinear spiral order for the AFM-2 phase, and a spin-canting cone-like structure, for the AFM-3 one. The magnetic structure of the low temperature phases, however, remains still unknown.

More recently, Yamasaki et al[4] reported a structural, dielectric, magnetic and polarization study done in Eu$_{1-x}$Y$_x$MnO$_3$ single crystals ($x$ = 0, 0.1, 0.2, 0.3 and 0.4). The temperature dependence of the electric polarization was obtained by measuring the pyroelectric current, after cooling the sample, provided with silver paste painted electrodes, under rather high electric fields (2kV/cm).[4] For the particular case of $x$ = 0.2, Yamasaki et al[4] reported the absence of a ferroelectric phase, and they attributed the discrepancy in the results reported in Refs. 1 and 4 to the non-stoichiometry of the samples.

The ferroelectric character of $Eu_{0.8}Y_{0.2}MnO_3$ below $T_{AFM-2}$ is then an open question. In most of the magnetically induced ferroelectrics, ferroelectricity is originated from a variety of spiral magnetic structures,[5] and can be explained in terms of the inverse Dzyaloshinski-Morya model, which states that the electric polarization is expected as:[6,7]

$$\vec{P} = \sum_{i,j} A\vec{e}_{ij} \times (\vec{S}_i \times \vec{S}_j), \qquad (1)$$

where $\vec{e}_{ij}$ denotes the unit vector connecting the interacting neighbour $\vec{S}_i$ and $\vec{S}_j$ spins, and A is the coupling constant between electric polarization and magnetic momenta, determined by both spin exchange interaction and spin-orbit coupling.[8] The origin of ferroelectricity in some rare-earth manganites and other magnetic ferroelectrics, such as $Ni_3V_2O_8$ [9] and $CoCrO_4$ [10] has been attributed to that mechanism. So, in the magnetic ferroelectric rare-earth manganites, the ferroelectricity has an improper character, having a completely different origin from that one in conventional ferroelectrics, leading, in general, to very small values of the spontaneous electric polarization.

The measurement of the electric polarization reported in the works referred to above was carried out using rather high electric fields, yielding, however, a very small saturation polarization value (P ≈ 30 – 50 nC/cm$^2$).[1,4] In $Eu_{0.8}Y_{0.2}MnO_3$, the existence of Y-impurities, with a smaller ionic radius than $Eu^{3+}$, can enhance the polarisable character of the crystal lattice. Thus, induced polarization is expected in this compound. In materials where small spontaneous electric polarization and an easily induced electric polarization coexist, understanding experimental data concerning the temperature dependence of the electric polarization, calculated from the time integration of the pyroelectric current, requires very careful analysis. These aspects are particularly relevant when, after cooling the sample under high electric fields, the pyroelectric currents are measured in heating runs. It is worth mentioning that the induced polarization, obtained with high enough poling electric fields, may completely hide the spontaneous contribution.

In this letter, we report a detailed study of the polar properties of $Eu_{0.8}Y_{0.2}MnO_3$ through measurements of the thermally stimulated depolarization current, pyroelectric current and polarization reversal (P(E)). As far as we know, a study of the temperature dependence of the P(E) relation has not yet been reported in current literature. Our aim is to clarify the controversial aspects of the phase sequence of $Eu_{0.8}Y_{0.2}MnO_3$ system, especially its low temperature ferroelectric properties.

High quality $Eu_{0.8}Y_{0.2}MnO_3$ ceramics were prepared by the sol-gel combustion method. A detailed study in EuMnO3 and GdMnO3 ceramics prepared in this way, has lead to results very similar to the ones obtained in the corresponding single crystals.[11] The phase purity and the crystallographic characterization of the ceramic samples were checked using X-ray powder diffraction and scanning electron microscopy. No secondary phases or significant deviation of the oxygen occupancy from the values expected for stoichiometric $Eu_{0.8}Y_{0.2}MnO_3$ were observed. A detailed dielectric, magnetic and magnetodielectric characterization of $Eu_{0.8}Y_{0.2}MnO_3$ ceramics will be published elsewhere.

Rectangular shape samples were prepared from the ceramic pellet, and gold electrodes were deposited using the evaporation method. The study of thermally stimulated depolarization currents was carried out in sequential thermal cycles as follows: (i) cooling the sample from 50K to 10K under a polarizing electric field $E_p$ up to 11.3 kV/cm; (ii) heating the sample, after short-circuiting it for 30 minutes, from 10K to 50K under zero electric field. The thermally stimulated depolarization currents were measured as a function of temperature, with a standard short-circuit method, using a Keithley electrometer, with 0.5 pA resolution, while keeping a fixed temperature rate.[12] The temperature dependence of the corresponding polarisation was obtained by the time integration of the current density. The sample temperature was measured with an accuracy better than 0.1 K. P(E) was recorded between 45 K and 10 K, using a modified Sawyer-Tower circuit[13], operating at 330 mHz.

Figure 1 shows the hysteresis loops obtained at several selected temperatures. Between 40K and 28K, a linear P(E) dependence is observed. As the temperature decreases from 27K towards 23K, hysteresis loops can be detected, with an elongated shape. A limited value of the saturation of the electric polarization could not be achieved, even for electric fields of up to 15 kV/cm, which shows how easily this material can be polarized. The most remarkable result is the retrieval of the linear P(E) relationship below $T_{AFM-3}$, which clearly reveals that ferroelectricity is intrinsic in the AFM-2 magnetic phase, and not in the AFM-3 phase, as previously suggested.[1] This feature provides a clear evidence for a ferroelectric character, only in the re-entrant AFM-2 phase.

The ease of inducing polarization in this material, revealed in the results referred to above, is corroborated by the experimental data obtained from the study of thermally stimulated depolarization currents, as we shall see below. Figure 2(a) shows the temperature dependence of the thermally stimulated depolarization current density J(T). In all measurements, J(T) curves show anomalies at $T_{AFM-2}$, and their amplitudes increase as the poling field increases. The temperature dependence of both real ($\varepsilon'_r$) and imaginary ($\varepsilon''_r$) parts of the dielectric constant, measured in a heating run at 100 kHz, is presented in Figure 2(b). $\varepsilon'_r(T)$ and $\varepsilon''_r(T)$ exhibit a pronounced anomaly at $T_{AFM-2}$, in good agreement with the reported results in single crystals.[1,2] A faint but clear anomaly in $\varepsilon''_r(T)$ at $T_{AFM-3}$, marks the AFM-2/AFM-3 phase transition. Figure 2(c) shows the temperature dependence of the electric polarization, which was obtained from time integration of the current density displayed in Figure 2(a). All the polarizations emerge below 30K and their saturation values are strongly dependent on the polarizing electric field. The saturation value of the polarization for $E_p$ = 2.8 kV/cm ($P_s \approx 1\times10^{-4}$ C/m$^2$) is lower but of the same order of magnitude as the one along the *a*-axis reported by Hemberger *et al*[1] for $Eu_{0.8}Y_{0.2}MnO_3$ single crystals ($P_a \approx 5\times10^{-4}$ C/m$^2$) under a poling electric field of 1 kV/cm. This difference should be associated with both random orientation of ceramic grains, and surface effects, which apparently reduce the actual polarization in granular samples. We should emphasize the ease of polarizing $Eu_{0.8}Y_{0.2}MnO_3$, even for rather small electric fields, of below 70 V/cm.

The electric current density referred to above is not all due to a spontaneous ferroelectric polarization. If it were, the corresponding polarization would have the same temperature behaviour as the remanent polarization, obtained from P(E) measurements. In this work, the measurement of pyroelectric current was carried out as follows: a low poling electric field ($E_p \approx 1$ V/cm) was applied inside the ferroelectric phase (T=25K). This procedure provided the alignment of the spontaneous electric dipoles, minimizing the excitation of other dipolar system. Afterwards, the sample was cooled to 10K, when the electric field was removed and the sample short-circuited for 30 minutes. The measurement of the electric current was then carried out in a heating run without applied electric field. Figure 3 shows the pyroelectric current density ($J_p$) as a function of the temperature. Anomalies are detected at both $T_{AFM-2}$ and at $T_{AFM-3}$, with opposite signs, yielding to a spontaneous electric polarization between $T_{AFM-3}$ and $T_{AFM-2}$. We calculated the spontaneous polarization ($P_s$) by time-integrating the pyroelectric current, presented in Figure 3. The analysis of the data displayed in Figure 1, allows us to determine the temperature dependence of the remanent polarization ($P_r$). Figure 4 shows the temperature dependence of both spontaneous and remanent polarizations, and despite the low values of $P_s$ and $P_r$, there is a good agreement between both results obtained using different techniques, in the 22K – 40K temperature range. As is general the case, the amplitude of the $P_r$ obtained from hysteresis loops is larger than that obtained from the analysis of the pyroelectric current. The polarization obtained by time integration of the pyroelectric current observed below 22K is an additional contribution, associated with an induced polarization, due to the poling electric field ($E_p \approx 1$ V/cm).

For comparison purposes, we have also measured the pyroelectric current in the sample studied in this work, provided with silver paste painted electrodes, and followed the experimental procedure referred just above. No anomalies could be found in the temperature dependence of the measured current below 50K.

To summarize, we have investigated the ferroelectric character of the low temperature magnetic phases of $Eu_{0.8}Y_{0.2}MnO_3$. The absence of both spontaneous and induced electric polarization in the AFM-1 phase corroborates its incommensurate collinear spin structure,[4] which prevents electric polarization from emerging, even after poling the samples with high electric fields. The existence of a spontaneous polarization in the AFM-2 phase provides evidence for spatial inversion loss, which is only recovered in the AFM-3 phase. Assuming that ferroelectric polarization is likely to be driven by the Mn 3d spin system arranged in a noncollinear cycloidal structure, a noncollinear to another spin arrangement, which prevents spontaneous electric polarization, will have to take place at $T_{AFM-3}$. Otherwise, both lower temperature magnetic phases allow for induced electric polarization. The collinear spin arrangement of the AFM-3 phase may become unstable under applied electric fields, via a strong spin-lattice coupling mechanism, giving rise to the observed induced polarization.[14] From the experimental results presented in this report, ferroelectricity and ferromagnetism do not coexist in the same phase; consequently, $Eu_{0.8}Y_{0.2}MnO_3$ cannot be a multiferroic material.

## ACKNOWLEDGMENTS

This work was supported by Fundação para a Ciência e Tecnologia, through the Project PTDC/CTM/67575/2006 and by Program Alβan, the European Union Program of High Level Scholarships for Latin America (scholarship no. E06D100894BR).

**Captions**

Figure 1. P(E) recorded at 330 mHz, for fixed different temperatures.

Figure 2. Temperature dependence of: (a) the thermally stimulated depolarization current density, measured in a heating run, after poling the sample under several electric fields; (b) the real ($\varepsilon'_r$) and imaginary ($\varepsilon''_r$) parts of the dielectric constant measured for 100 kHz; (c) the electric polarization, obtained from time integration of the thermally stimulated depolarization current density. Inset to (c) shows an expanded view of the induced polarization, after poling the sample under 10 V/cm and 70 V/cm.

Figure 3. Temperature dependence of the pyroelectric current density, measured in a heating run, after cooling the sample with an applied electric field 1 V/cm.

Figure 4. Temperature dependence of the spontaneous polarization obtained from the time integration of the pyroelectric current, and of the remanent polarization, obtained from the P(E).

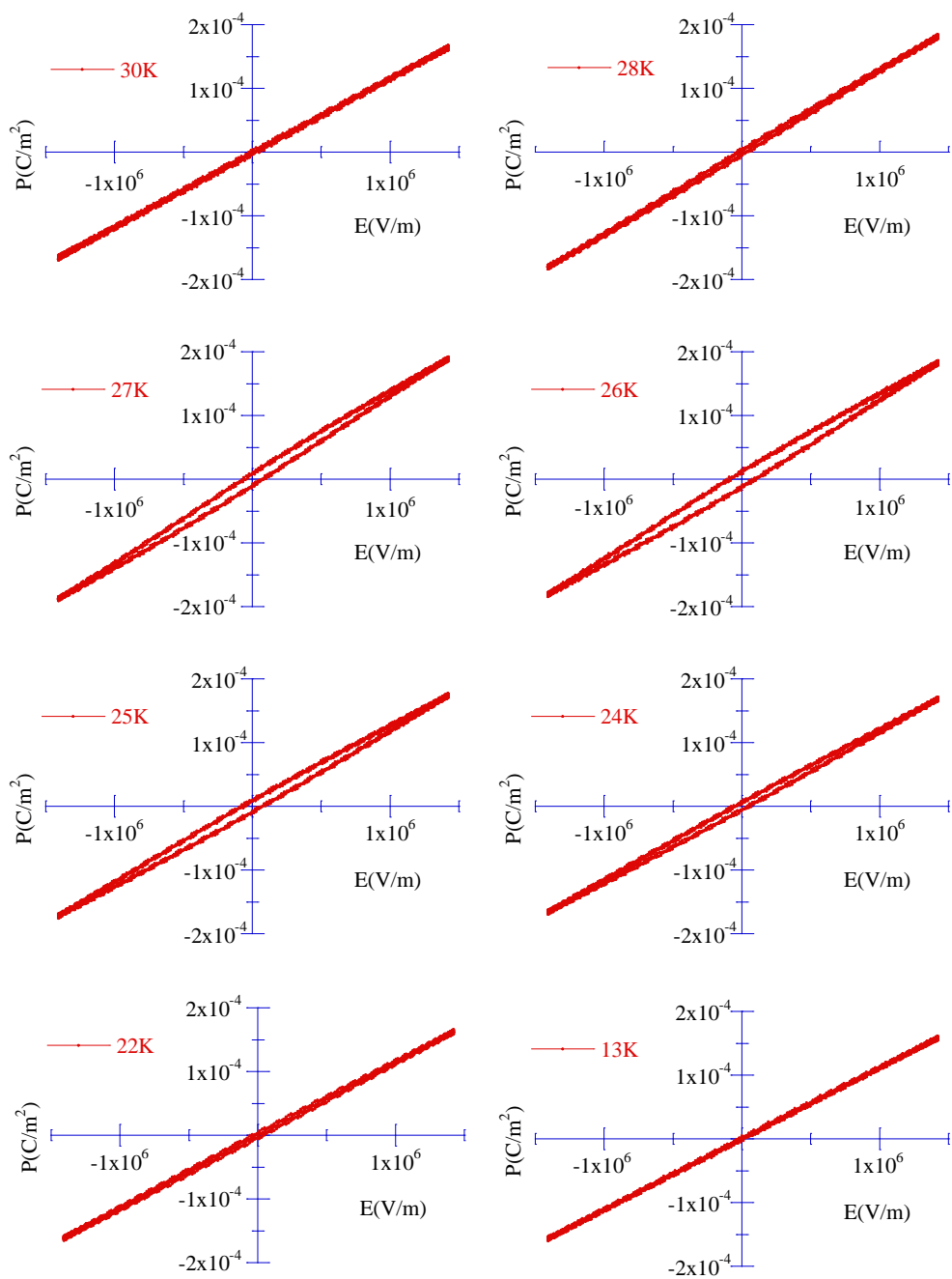

Figure 1

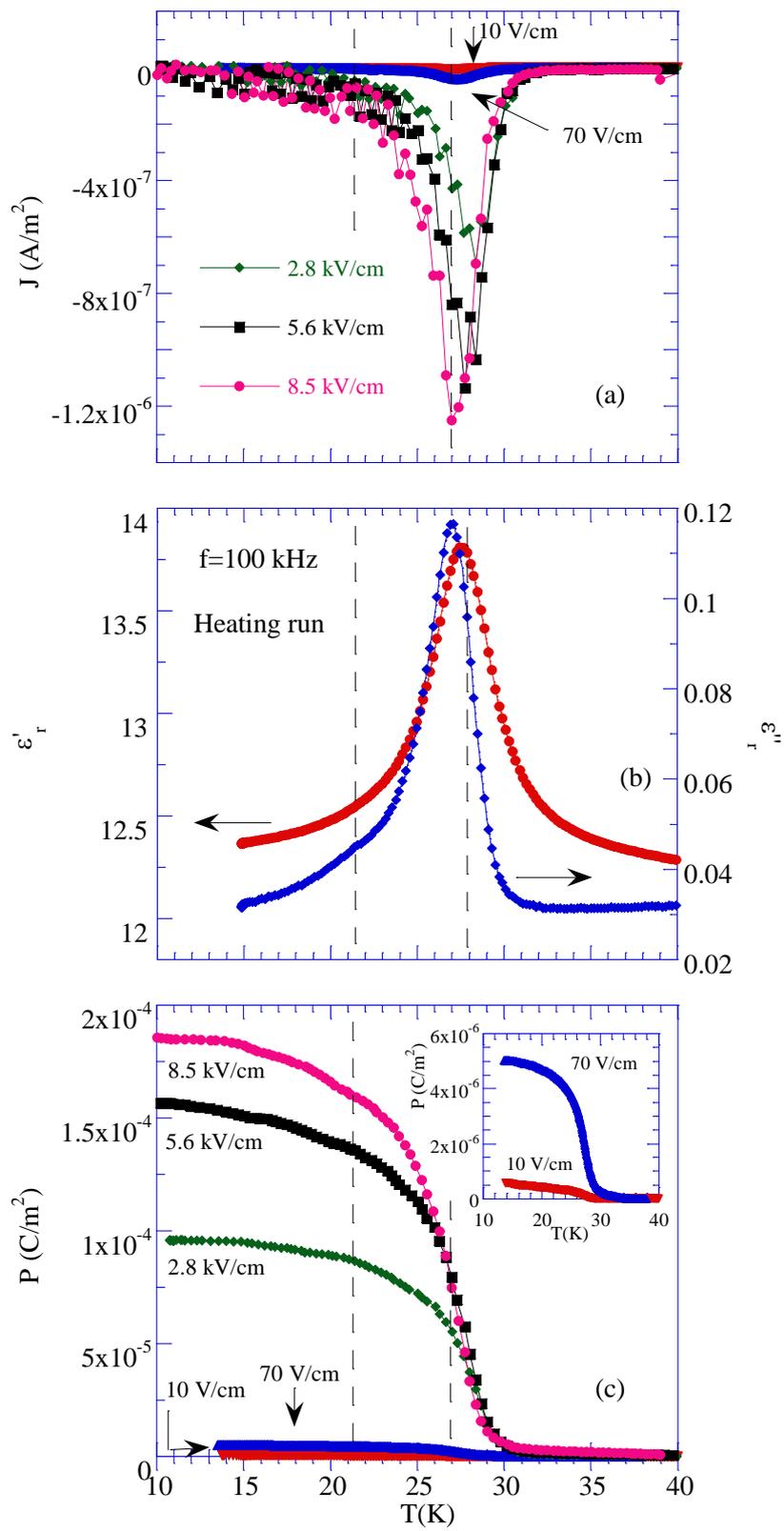

Figure 2

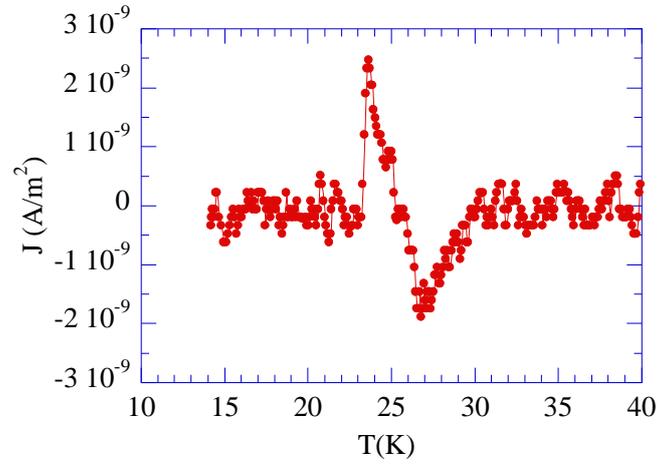

Figure 3

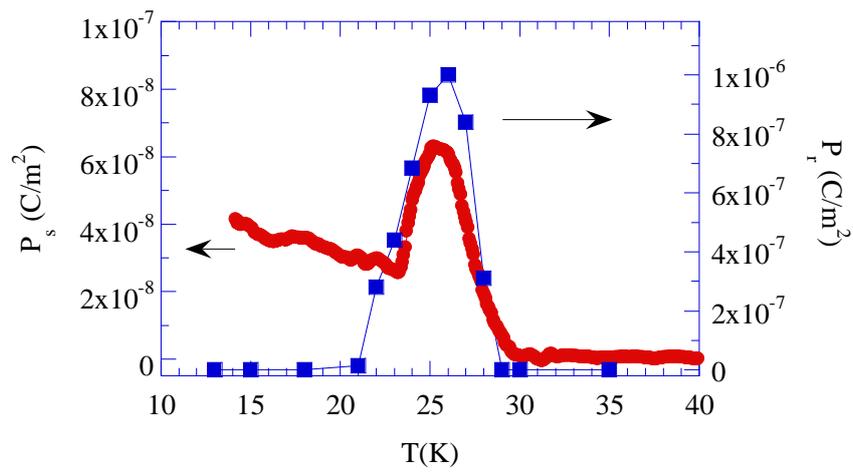

Figure 4